# A retrospective on the 2025 Atlantic hurricane season

**Charles W. Powell**
*ICCS & DAMTP, University of Cambridge, UK*

## Introduction

Tropical cyclone (TC) activity is strongly controlled by sea surface temperatures (SSTs); temperatures exceeding 26.5°C are needed for TC formation (e.g. McTaggart-Cowan *et al.*, 2015). However, other environmental conditions are required to support the formation of tropical depressions (cyclogenesis), their organisation into coherent deep circulations and intensification to reach tropical storm strength (McTaggart-Cowan *et al.*, 2015; Rajasree *et al.*, 2023). Factors promoting the formation of TCs include upper-level divergence, which promotes surface convergence (Gray, 1968) and aids the development of 'seed disturbances' from which most TCs grow (Sears and Velden, 2014). Cyclogenesis is also favoured by factors that promote deep convection such as large-scale instability and requires the formation of a large column of nearly saturated air (i.e. large relative humidity (RH), particularly at mid-levels) that allows disturbances to grow efficiently (Emanuel, 2003). Other factors can inhibit TC formation, such as vertical wind shear, which disrupts the organisation of tropical depressions by displacing the low- and upper-level circulation centres and promotes ventilation of dry air into developing TCs (Rios-Berrios *et al.*, 2024a). Similarly, developing convective systems are suppressed in regions with dry mid-level air and large-scale subsidence that increases stability and disrupts low-level convergence. TC activity is therefore influenced by variability in the aforementioned factors.

In the North Atlantic basin, the largest source of interannual variability in TC activity is the El-Niño Southern Oscillation (ENSO). El Niño conditions increase vertical wind shear and subsidence in the Atlantic basin, suppressing TC activity, whilst La Niña conditions tend to favour active seasons via a weakening of upper-level winds (Gray, 1984; Klotzbach, 2011). Whilst some TCs form *in situ*, particularly early and late in the season in the Gulf of Mexico and Caribbean Sea (Corporal-Lodangco *et al.*, 2014), the majority of TCs in the North Atlantic form in the main development region (MDR) from seed disturbances called African easterly waves (AEWs; Russell *et al.* (2017)) that arise from instability of the mid-level African easterly jet. Factors influencing the frequency and strength of AEWs, either by modifying the strength of the African easterly jet – such as the West African Monsoon (WAM) – or modifying the environmental conditions that support the development of AEWs, represent another important source of variability in TC activity. It has been shown that convectively coupled Kelvin waves (KWs) can influence TC activity by modulating the large-scale environment and promoting the evolution of AEWs (e.g. Schreck, 2016; Lawton *et al.*, 2022), affecting the likelihood of cyclogenesis. The development of AEWs is influenced by other environmental factors: deep convection can be suppressed by dry air from the Sahara desert that is carried into the Atlantic, and the warm SSTs needed to drive cyclogenesis are influenced by small-scale SST patterns that are coupled with low-level trade winds and circulation patterns (Kim *et al.*, 2023). Finally, the Madden–Julian Oscillation (MJO) is the leading mode of intraseasonal variability in the tropics globally, which influences TC activity throughout the Atlantic basin (Maloney and Hartmann, 2000; Klotzbach and Oliver, 2015). Although the eastward-propagating convective patterns associated with the MJO are confined to the western Pacific, upper-tropospheric wind anomalies propagate around the globe that can temporarily enhance upper-level divergence over the North Atlantic basin, favouring cyclogenesis (Barrett and Leslie, 2009).

In this paper, we review the factors that influenced activity during the 2025 North Atlantic hurricane season. The 2025 season is an interesting case study since common sources of variability such as ENSO and the MJO were weak, allowing other factors to more strongly drive changes in TC activity. ENSO conditions transitioned from neutral to weak La Niña during the season (NOAA, 2025), and the number of days with favourable MJO conditions in the North Atlantic during the Atlantic season (June to November) was the second lowest since 1991 according to the RMM index (Wheeler and Hendon, 2004). The 2025 season also saw the fifth warmest seasonal-mean daily SSTs in the MDR (using the OISST dataset; Huang *et al.*, 2021) yet ended with only slightly above-average activity (using HURDAT2; Landsea and Franklin, 2013). In an average Atlantic hurricane season, there are 14 named storms, of which seven become hurricanes and three reach major hurricane status (NHC, 2026). In 2025, there were 13 named storms, of which five became hurricanes. It is therefore valuable to examine why 2025 was comparatively less active. Finally, it is noteworthy that most activity occurred in just four storms that reached major hurricane status, especially considering that three of the five hurricanes in 2025 reached Category 5, the most since 2005 – an extremely active season. Although projections of TC activity in the changing climate are uncertain, the 2025 season appears to fit the model-predicted trend of decreasing overall activity alongside increasing intensity of individual TCs (Knutson *et al.*, 2020).

In this study, we identify which of the common drivers of TC activity in a typical season were more or less influential in the 2025 season, identify other factors that were more influential in 2025 and seek to explain the relative lack of activity in the context of above-average SSTs. As a single-season analysis, this study cannot directly attribute TC activity to individual drivers. Instead, we identify differences between periods with a relative change in activity within the season. Overall, reviewing the driving factors of a single season in this way is valuable as a case study of theoretical or statistical relationships identified in the literature, such as the influence of KWs on AEW development. First, we summarise the 2025 North Atlantic hurricane season, review the state of the dominant drivers (SSTs and ENSO), and identify the unusual qualities of the 2025 season. We then address the large-scale environmental conditions in the North Atlantic. Using a wave-filtering analysis, we explore AEWs in the 2025 season and, in particular, the influence of KWs and favourable MJO phases. Finally, we examine the contribution of SST and low-level circulation patterns to intraseasonal variability in 2025.

## Season overview

Figure 1 shows TC tracks from the 2025 season and regional boxes used in later analysis. Climatologically, the Atlantic hurricane season lasts from 1 June to 30 November, with peak activity during August and September. In 2025, activity began on 23 June with three early short-lived tropical storms in the east of the basin but otherwise little activity throughout July and early August. Approaching the climatological peak, activity restarted when Hurricane *Erin* formed on 11 August and went on to reach Category 5 strength. Thereafter, TC activity ceased throughout much of the peak season, similar to the lull observed in 2024 (Klotzbach *et al*., 2025). Two late-season clusters of activity brought cumulative season activity to above-average levels, again similar to the back-loaded 2024 season. In mid-late September, Hurricanes *Gabrielle*, *Humberto* and *Imelda* formed, though none made landfall. In late October, Hurricane *Melissa* became the third CCategory 5 hurricane of the season and the joint-strongest landfalling Atlantic hurricane with a maximum sustained wind speed of 185mph and central pressure of 892hPa at landfall. Post-season analysis has verified a wind gust reading of 252mph via dropsonde (UCAR, 2025), setting a new global record for the strongest dropsonde-measured wind gust in a TC.

SSTs are the strongest indicator of TC activity; the record-breaking 2005 season (ACE ~ 240) was widely attributed to (at the time) record-warm SSTs (e.g. Trenberth and Shea (2006)) and a long-term increase in TC activity since 1994 has been attributed to warmer tropical SSTs (e.g. Goldenberg *et al*., 2001; Emanuel, 2005; Klotzbach and Gray, 2008). The 2023 and 2024 Atlantic hurricane seasons saw record high SSTs throughout the Atlantic basin, surpassing 2005 – see Figure 2, which shows daily mean SSTs in the MDR from 1991 to 2020, with 2023, 2024 and 2025 highlighted. Despite the unprecedented temperatures, the 2024 season was remarkably quiet during the peak season (Klotzbach *et al*., 2025) owing to a northward shift in the track of AEWs, as well as strong vertical wind shear and broad-scale subsidence. Nonetheless, the season ended with above-average accumulated cyclone energy (ACE) owing to a late-season cluster of activity including two major hurricanes. Activity in the 2023 season was near-average, as El Niño conditions suppressed cyclogenesis. In 2025, MDR SSTs remained well above the long-term average. Daily mean MDR SST anomalies (relative to 1991–2020 climatology) averaged 0.47°C during the 2025 hurricane season, making it the fifth warmest on record. Alongside warm SSTs, five of the past six seasons have seen La Niña conditions during the peak season according to the relative Oceanic Niño Index (NOAA Climate Prediction Center, 2026b), although the La Niña signal was weak until late in the 2025 season.

The 2025 season was unusual in the sense that activity was intermittent, with extended quiet periods separated by just three clusters of activity that featured particularly intense TCs. The total number of storms was close to average, the number of hurricanes below average, but the number of major hurricanes above average. The season ended with a total ACE of 132.5 ($10^4$kn$^2$); a measure of activity that combines the strength and duration of storms. Figure 3 shows the cumulative ACE in the 2025 season compared to 1991–2020 climatology, alongside a timeline of the season, clearly showing the three periods in which most of the cumulative ACE was generated. Comparison seasons (discussed below) are also shown. The 2025 season is classified as 'above-normal' according to the NOAA definition (based on % of the 1951–2020 median), consistent with the number of strong TCs observed. However, the classification is contradicted by the near-average number of storms and long periods of no/weak activity. The intermittency of the season is demonstrated by the fact that around 85% of the total season ACE occurred in just four out of 13 storms. As noted earlier, this appears consistent with projected trends in TC activity as the climate warms (Knutson *et al*., 2020).

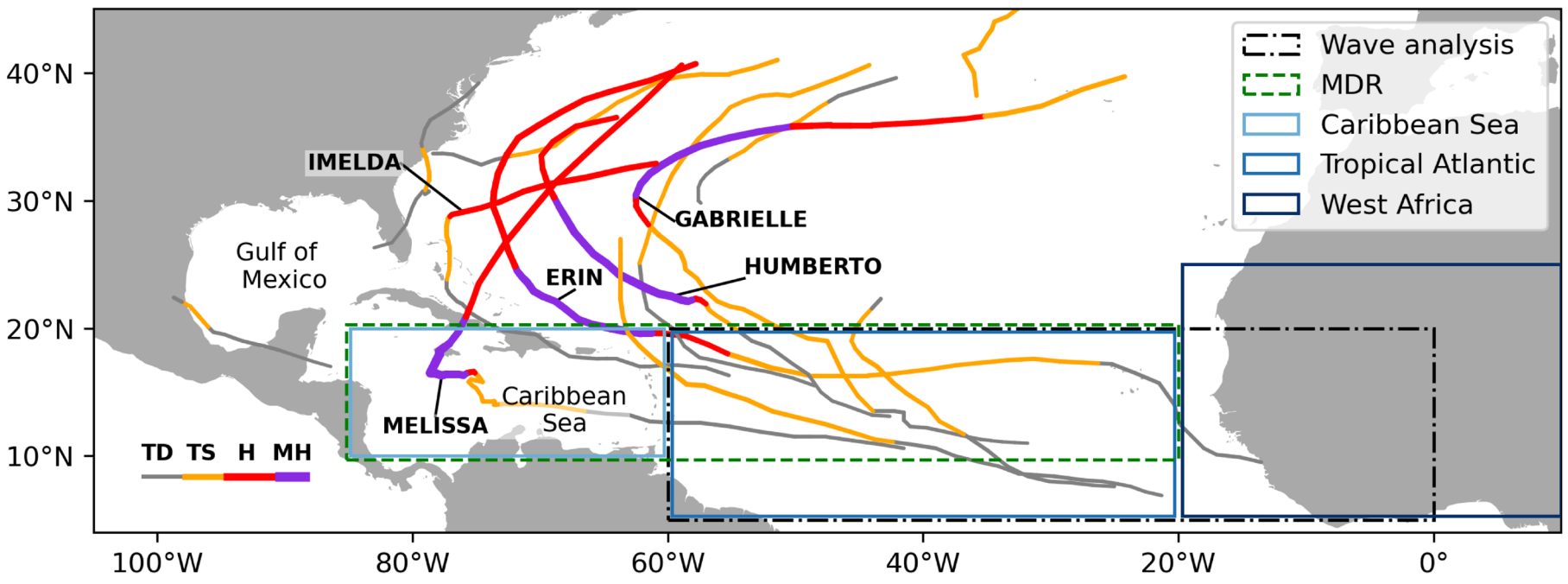

*Figure 1. Summary map of the Atlantic basin, 2025 TC tracks and region definitions used in this paper. Tracks are coloured by Category (tropical depression TD; tropical storm TS; hurricane H; major hurricane MH) with line width proportional to maximum sustained wind speed. Hurricanes labelled. Coloured boxes indicate the MDR used in Figure 2, wave analysis region used in Figures 5 and 6, and the Caribbean Sea, Tropical Atlantic and West Africa regions used in Figure 6. Track data for 2025 obtained from ATCF realtime (Sampson and Schrader, 2000).*

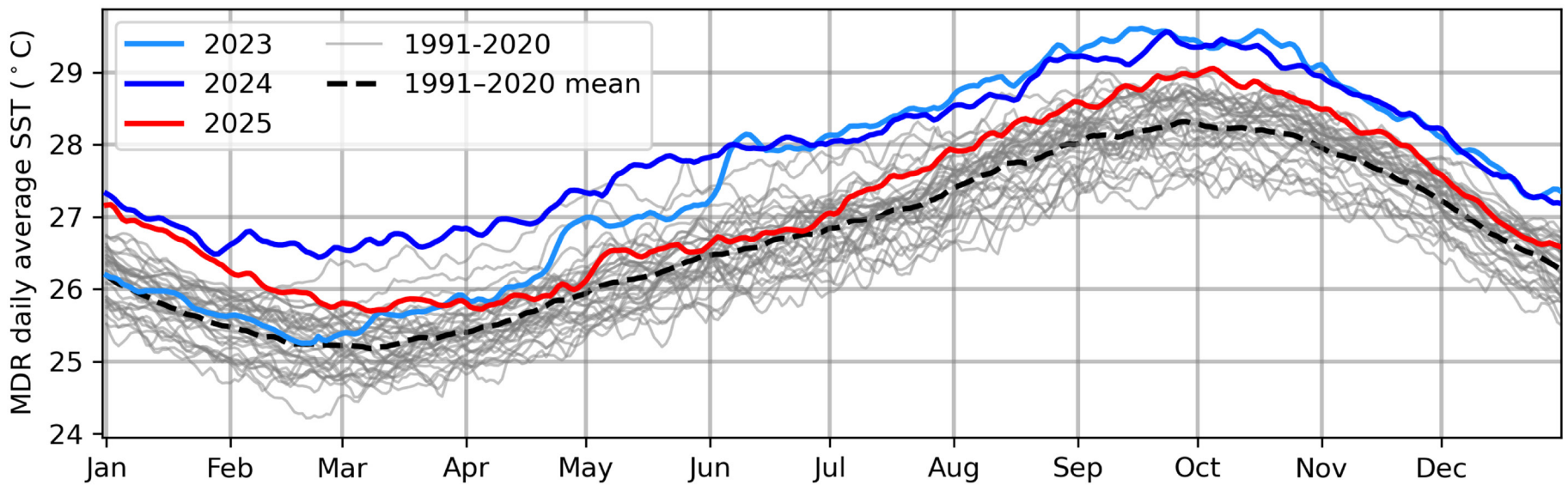

*Figure 2. Daily mean SSTs (°C) in the Atlantic MDR (10°–20°N, 20°–85°W) in 2023–2025 (coloured), compared with 1991–2020 climatology (grey, mean in dashed black). Data obtained from NOAA OISSTv2.1 (Huang* et al.*, 2021).*

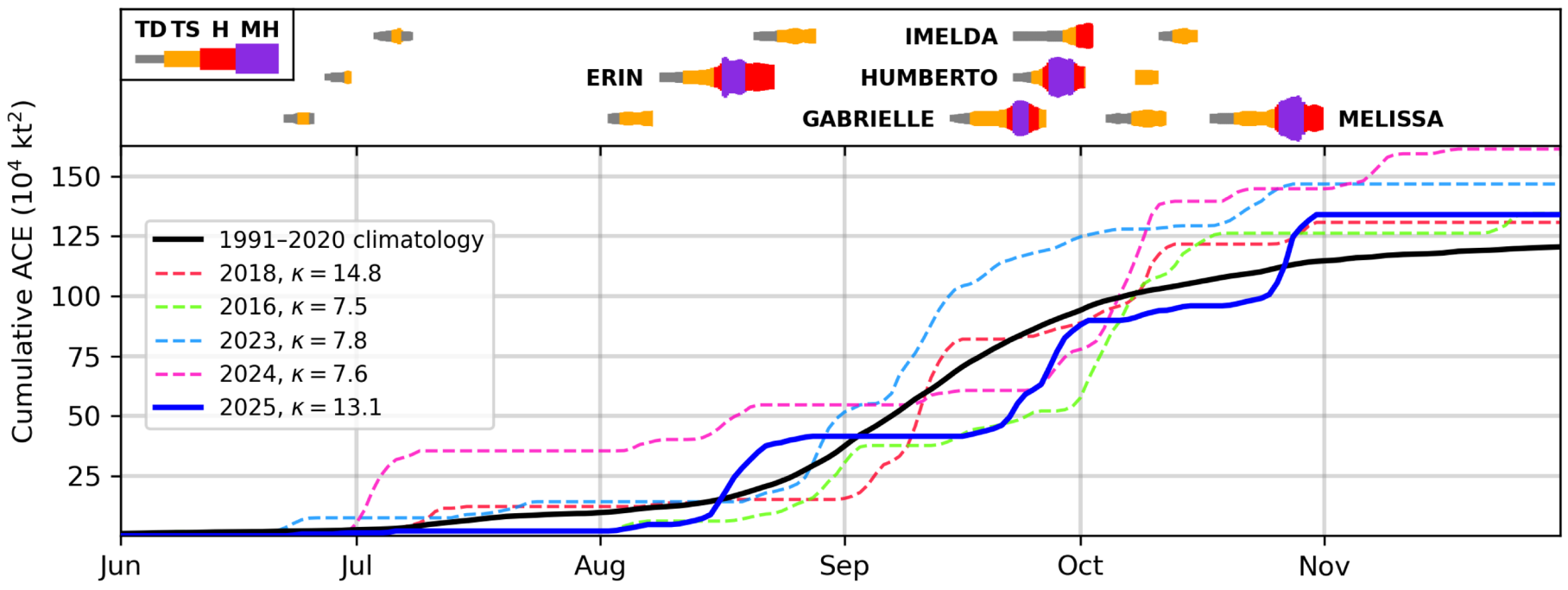

*Figure 3. Cumulative accumulated cyclone energy (ACE) in 2025 compared with the long-term (1991–2020) average and specific comparison years referred to in the text. Lifecycles of 2025 TCs inset, coloured as in Figure 1, with hurricanes labelled. Data for 2025 obtained from ATCF realtime (Sampson and Schrader, 2000) and for 1991–2020 from HURDAT2 (Landsea and Franklin, 2013).*

Another measure of intermittency is the kurtosis $\kappa$ of the daily ACE distribution (using HURDAT2; Landsea and Franklin, 2013),

which measures the tailedness of the distribution (Wilks, 2006): increasing values correspond to increasingly extreme deviations from the mean. In the modern climatology period from 1991 onwards, the 2025 season is uniquely placed as an intermittent season with both near-normal ACE (within 20% of the mean ~120) and warm MDR SSTs. The 2025 ACE distribution has the second-highest kurtosis ($\kappa \approx 13$) in this period (with mean $\kappa \approx 7$), exceeded only by the 2018 season ($\kappa \approx 15$). However, MDR SSTs in 2018 were below average (−0.28°C), with neutral ENSO conditions developing into El Niño late in the season, and a strong WAM (NOAA, 2026). The intermittency in 2018 can therefore be attributed to the frequent but relatively weak activity amid otherwise unfavourable conditions for TC activity, punctuated by a Category 5 and a long-lived Category 4 hurricane which generated extreme ACE values compared to the rest of the season, increasing $\kappa$. The near-normal ACE in 2018 exceeded expectations based on SSTs, whereas in 2025, the near-normal ACE can be viewed as unusual based on SSTs. Only the 2016 season has comparable MDR SSTs to 2025 (0.39°C) and near-normal ACE (~140) in the modern climatology period, but with near-average kurtosis ($\kappa \approx 8$) indicating steady accumulation of ACE throughout the season amid favourable La Niña conditions. All other seasons with similar or greater SSTs (2005, 2010, 2017, 2020, 2023, 2024) saw well above-average activity, with the exception of 2023 which featured a strong El Niño. In the following sections, we identify the unfavourable conditions that led to the relative lack of TC activity in 2025 and discuss factors that contributed to the strong intraseasonal variability compared to recent climatology, manifested as clusters of very active days interspersed with extended inactive periods.

## Large-scale environmental conditions

The large-scale thermodynamic environment in the Atlantic basin in 2025 was generally unfavourable for TC activity. Figure 4 shows anomalies of four key environmental factors, averaged over the quiet early season (June and July) and the more active peak season (August–October). The latter period includes the peak-season lull from mid-August to mid-September, though the remainder of the period was more active. We consider sea-level pressure (SLP), relative SSTs (relative to the mean SST in the tropical belt 20°N–S), 200–850hPa temperature lapse rate, and mid-level RH at 500hPa. These factors act to favour or suppress cyclogenesis by providing or disrupting large-scale support for convection. Note that Figure 4 shows a large-scale and time-averaged picture of the environment; localised support for cyclogenesis can – and did – drive convection over short periods. Moreover, there is a two-way feedback between TCs and their surrounding environment: TCs are associated with low SLP, extended regions of moist mid-level air and tend to leave a cool SST wake in their path (Karnauskas *et al*., 2021). To account for contamination of the monthly average by strong long-lived TCs, we remove days with daily ACE exceeding 1 unit (0 days in June–July, 28 days in August–October).

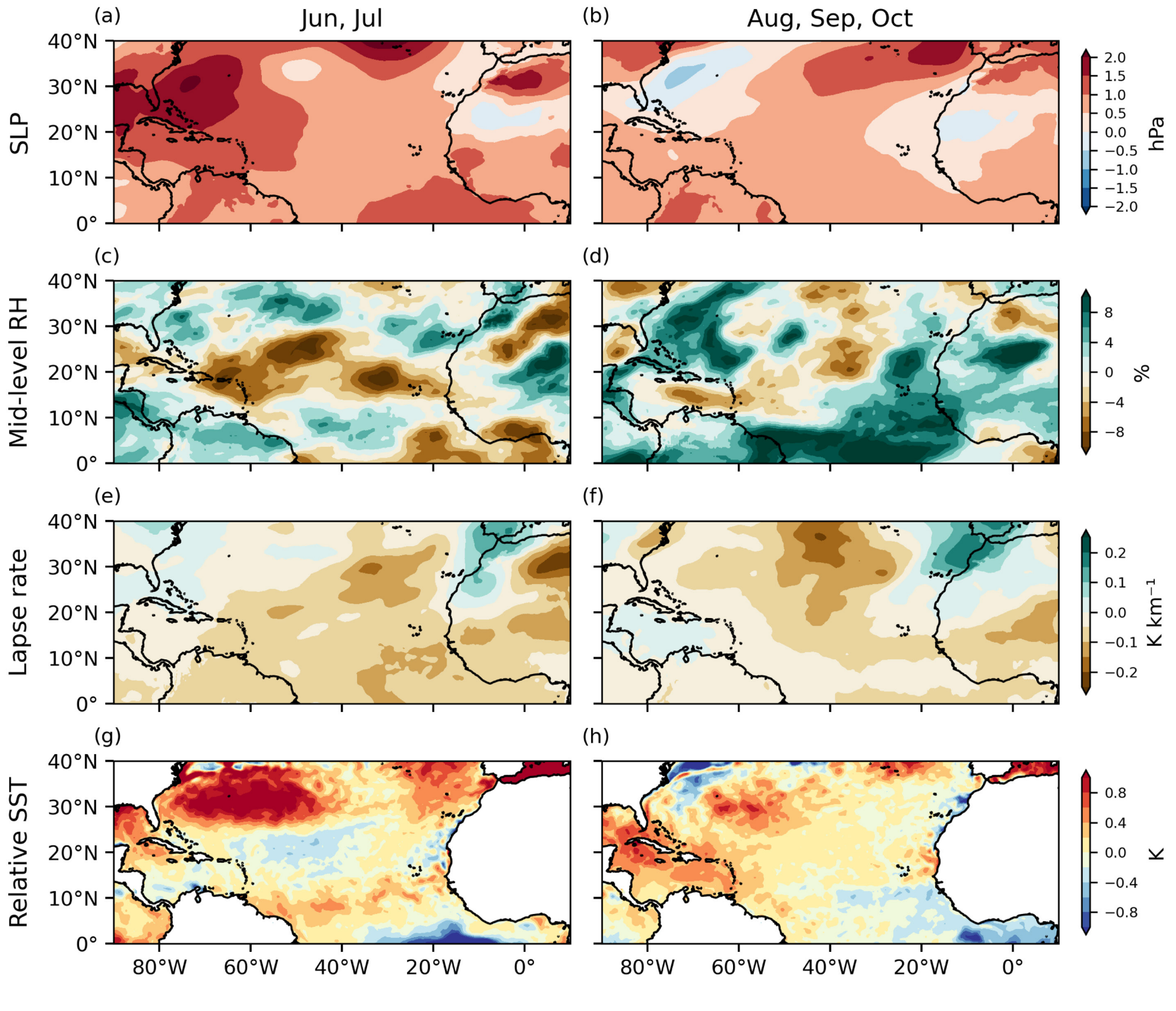

*Figure 4. Monthly-mean anomaly (relative to 1991–2020 climatology) in June–July and August–October 2025 of (a, b) sea-level pressure, (c, d) 500hPa relative humidity, (e, f) 850–200hPa temperature lapse rate and (g, h) relative sea-surface temperature. Days with ACE > 1 are removed from the averages. Data from preliminary ERA5 reanalysis (Hersbach* et al., *2020).*

The 2025 season saw basin-wide high SLP throughout. The typical pressure pattern in the Atlantic basin during the summer months features a subtropical high pressure system that drives the easterly trade winds in the tropics. Anomalous high summertime SLP (Figure 4a, b) has been linked to suppressed TC activity owing to increased large-scale subsidence of air that dries out the mid-troposphere, as well as indicating the presence of a persistent tropical upper-tropospheric trough (TUTT) that increases vertical wind shear (see TUTT discussion in 'Circulation and SST Patterns'; Knaff, 1997). High mid-level RH (Figure 4c, d) is essential to support deep convection; a dry mid-troposphere therefore reduces the likelihood of cyclogenesis, acting as a positive feedback loop that maintains the SLP anomaly. Although the dry anomalies are smaller scale than the basin-wide high SLP, they are focussed in the tropical Atlantic near the coast of West Africa in June and July. These anomalies are also indicative of the dry Saharan air layer (SAL) extending into the Atlantic, disrupting the development of AEWs. Examining differences between June, July and August (not shown) shows a wet/dry dipole that moves northwards from the Guinea Coast (southern coast of West Africa) in June to the Sahel in central West Africa in July and August, indicating a gradual onset of the WAM that usually occurs in late June (e.g. Vellinga *et al*. (2013)). The position and strength of the convectively active region of the WAM is a crucial factor in the generation of AEWs (Hsieh and Cook, 2005). The weak WAM and high SLP both result in weak AEWs that are less likely to form TCs during July and early August. Although the high SLP anomaly persists in August–October, the anomaly is weaker and lower SLP is present near the US west coast. Correspondingly, the tropical Atlantic mid-troposphere is generally wetter than climatology during August–October, indicating a less hostile environment for AEWs and TC development.

Another persistent feature of the 2025 season is low temperature lapse rates (Figure 4e, f), indicating a more stable atmosphere that is less favourable for deep convection. This is consistent with climate model predictions that lapse rates will decrease as the upper troposphere warms faster than the lower troposphere (e.g. Keil *et al*., 2021), resulting

RMetS Royal Meteorological Society

in increasingly hostile thermodynamic conditions for TC development and intensification. Although the troposphere was more stable throughout the season (particularly over West Africa, likely disrupting the WAM), the tropical Atlantic and Caribbean Sea became less stable in August–October, providing better support for the development of TCs. The spatial pattern of SSTs throughout the 2025 season also acted against the development of convective systems in the Atlantic basin. This is best understood using relative SST anomalies (Figure 4g, h), with the mean SST anomaly in the tropical belt (20°S to 20°N) subtracted. Relative SSTs more clearly indicate patterns that can drive convection since warm SSTs are less effective when the surroundings are also warm (e.g. Williams *et al.*, 2023). Although SSTs remained close to record warmth in the MDR (Figure 2), large parts of the Atlantic basin were also much warmer than average. In particular, the highest relative SST anomalies were found in the subtropical Atlantic, around 15°N. In a typical season, the Hadley cell circulation shifts slightly north of the equator during the summer months, with air rising in the tropics along the inter-tropical convergence zone (ITCZ) and sinking in the subtropics. Warmer SSTs in the subtropics are indicative of a weaker Hadley cell circulation that is less supportive of convection in the tropical Atlantic. The relative SST contrast between the tropics and subtropics was weaker in August–October compared to June and July. Other SST and circulation patterns played a role in the observed variability in activity during these months, such as the marked increase in SSTs in the Caribbean Sea that created a potent environment for Category 5 *Melissa* to form and rapidly intensify in October (see later discussion).

## Influence of Kelvin waves, MJO and WAM on African easterly waves

Russell *et al.* (2017) showed that 71% of North Atlantic TCs are linked to AEWs. Recent work has shown that KWs can support cyclogenesis by preconditioning the environment around AEWs (Lawton *et al.*, 2022; Lawton and Majumdar, 2023), and the convectively active phase of a KW has been shown to modulate synoptic-scale conditions that favour the development of TCs from AEWs in the days following its passage (Rios-Berrios *et al.*, 2024b). The MJO has also been shown to favour cyclogenesis during periods of anomalous upper-level divergence (Barrett and Leslie, 2009). We emphasise that KWs and the MJO do not directly initiate cyclogenesis, rather they influence AEW development, thereby increasing the likelihood of cyclogenesis. Here, we analyse AEWs in the 2025 Atlantic season using the lower troposphere (700hPa) relative vorticity anomaly $\zeta'$ (e.g. Jonville *et al.*, 2025b). AEWs are identified as positive $\zeta'$ values following a bandpass filter with period 2–6 days (e.g. Hopsch *et al.*, 2010) and wavenumbers 6–20 (e.g. Schreck *et al.*, 2012). To explore the role of MJO favourable periods, we use a bandpass filter with period 30–96 days and wavenumbers 1–5 applied to the velocity potential anomaly at 200hPa, $\chi_{200}$, with negative values corresponding to divergence in the upper troposphere. Similarly, the role of KWs is investigated using a bandpass filter with period 2.5–20 days, wavenumbers 1–14 and equivalent depths 8–90m applied to the outgoing long-wave radiation (OLR) anomaly; negative values correspond to convectively active KW phases (Ventrice *et al.*, 2012b). A 40-day buffer period is added to the beginning and end of the time period of interest (June to November) to reduce spectral leakage. Favourable MJO conditions and KW phases are identified where the wave-filtered signal is less than one standard deviation below the mean (both calculated within the wave analysis region).

Figure 5(a) shows a Hovmöller plot (longitude *vs* time) of the AEW-filtered, latitudinally averaged $\zeta'$ field in the wave analysis region (0°–60°W, 5°–20°N) indicated in Figure 1. TC tracks are overlaid whilst the TC remains in the wave analysis region and coloured by intensity as in Figure 1. Convectively active KW phases and favourable MJO conditions are indicated by blue and magenta shading, respectively. To assess the varying strength of AEWs through the season, Figure 5(b) shows the maximum relative vorticity anomaly $\zeta'_{max}$ along each AEW track compared to climatology, plotted at the time that the track passes 20°W, corresponding to the approximate time that the AEW crosses the West African coast into the Atlantic. AEWs that reach their maximum amplitude up to 4 days after the passage of a convectively active KW phase are highlighted in solid blue. The choice of lead-time is based on analysis of the time lag between cyclogenesis and KW peaks by Rios-Berrios *et al*. (2024b) and Lawton and Majumdar (2023). AEWs that experience the passage of a KW in their lifetime are highlighted in light blue. AEWs that produce TCs are circled in red, using post-season analyses (NHC, 2026) for confirmation.

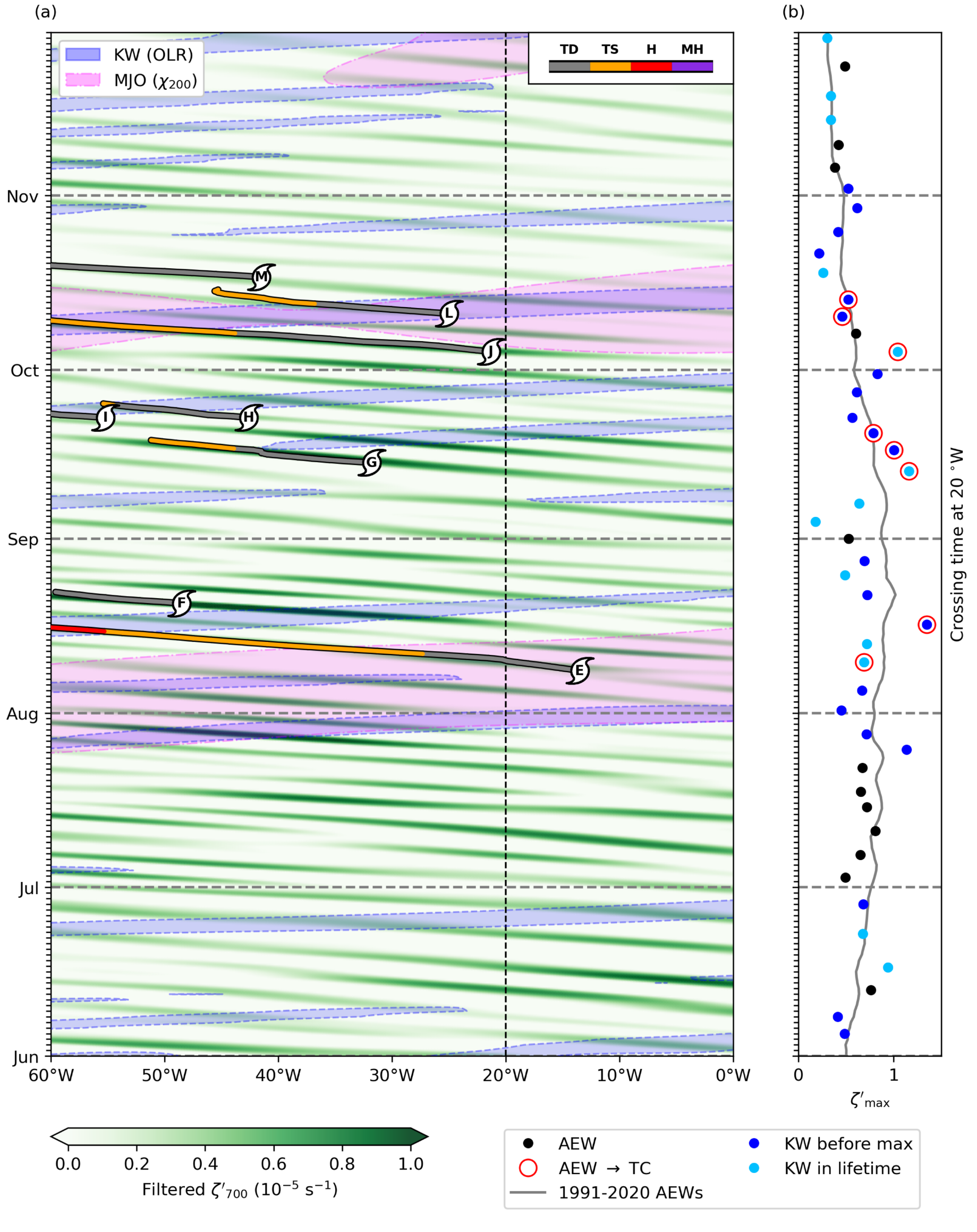

*Figure 5. (a) Hovmöller plot of the AEW-filtered relative vorticity anomaly $\zeta'$ (relative to 1991–2020 climatology) at 700hPa averaged over 5°–20°N latitude. Only positive values are shown for clarity. Favourable KW (MJO) signals are shown where the filtered OLR anomaly (velocity potential anomaly at 200hPa) is more than one standard deviation below the mean. TCs are overlaid whilst within the wave analysis region indicated in Figure 1. (b) AEW maximum amplitude (maximum relative vorticity anomaly $\zeta'_{max}$) versus crossing-time at 20°N. AEWs whose maximum amplitude occurs up to 4 days after a passing favourable KW signal are coloured dark blue, otherwise light blue if influenced by a KW during its lifetime. AEWs that form TCs are emphasised with a red circle. Data from preliminary ERA5 reanalysis (Hersbach* et al.*, 2020) other than $\chi_{200}$: 2025 data from CFS (Schneider* et al.*, 2013), climatology from NOAA PSL.*

Whilst the number of AEWs crossing 20°W during the 2025 hurricane season (46) is close to the climatological average (43) when calculated using our method, the maximum amplitude of the AEWs is below average for extended periods that coincide with low TC activity in the Atlantic basin, particularly July and most of August. Whilst weaker-than-average AEWs can still develop into TCs (e.g. *Erin* in mid-August), they are more likely to be disrupted during their evolution, reducing the likelihood of cyclogenesis (Agudelo *et al*., 2011). Six of eight AEWs that produce TCs in 2025 are close to or stronger than the climatological average. The frequency of AEWs entering the Atlantic basin appears uniform throughout the season, but owing to the disrupted WAM (due to anomalous subsidence and increased stability) and dry air over West Africa noted earlier and shown below, AEWs were generally weaker and consequently less likely to produce TCs during the first half of the season.

Figure 5(b) shows that many TC-producing AEWs are influenced by KWs shortly before reaching their maximum amplitude, consistent with Rios-Berrios *et al*. (2024b), and periods with a lack of KW activity (e.g. July and mid-August to mid-September) coincide with periods with weak AEWs. The latter observation may have played a role in the relative lack of TC activity during these periods – Thorncroft and Hodges (2001) showed that variability in the amplitude of AEWs, not just the number of AEWs, can influence TC activity. Over the season, the influence of KWs was consistent with climatology: 74% of AEWs during the 2025 season were influenced by KWs at some point in their lifetime, consistent with results from Lawton and Majumdar (2023) that 76% of AEWs were KW-influenced in a 39-year record from 1981 to 2019.

The suppressed development of AEWs in June–August can also be linked to episodes of strong vertical shear, dry air and increased subsidence over the tropical Atlantic, preventing development of AEWs downstream of West Africa. Figure 6(a) shows a timeseries of daily ACE highlighted where favourable MJO conditions and KW phases are present in the wave analysis region used in Figure 5. Figure 6(b) shows the 850–200hPa deep-layer vertical wind shear anomaly, regionally averaged in the Caribbean Sea, tropical Atlantic and West Africa (as shown in Figure 1). Thresholding of the 700hPa relative vorticity anomaly shown in Figure 5 was used to remove TCs and AEWs from the averages. Figure 6(b) shows that strong vertical shear was present for extended periods up to the middle of September in West Africa and the tropical Atlantic, disrupting the development of AEWs.

Figure 6(c) shows a WAM index defined as the standardised difference between the 925hPa wind speed and the 200hPa zonal wind speed averaged over 20°W–20°E, 3°–13°N, following Ndiaye *et al*. (2009). The WAM index measures the strength of the monsoon circulation: the monsoon can be considered 'active' when the index is positive, though onset of the WAM is difficult to define objectively (Fitzpatrick *et al*., 2015). AEW amplitudes typically decrease as the WAM weakens and dissipates in October, but the African easterly jet persists into November, continuing to produce weak AEWs (Afiesimama, 2007). Note that the circulation weakens in September whilst monsoon rainfall persists (Pu and Cook, 2012; Zhang and Cook, 2014), reducing the WAM index from September onwards as seen in the 1991–2020 climatology in Figure 6(c), despite continued influence on AEWs. The WAM index in Figure 6(c)is consistent with a disrupted WAM during the first half of the season (JJA) noted earlier, as well as a stronger WAM during late September and October.

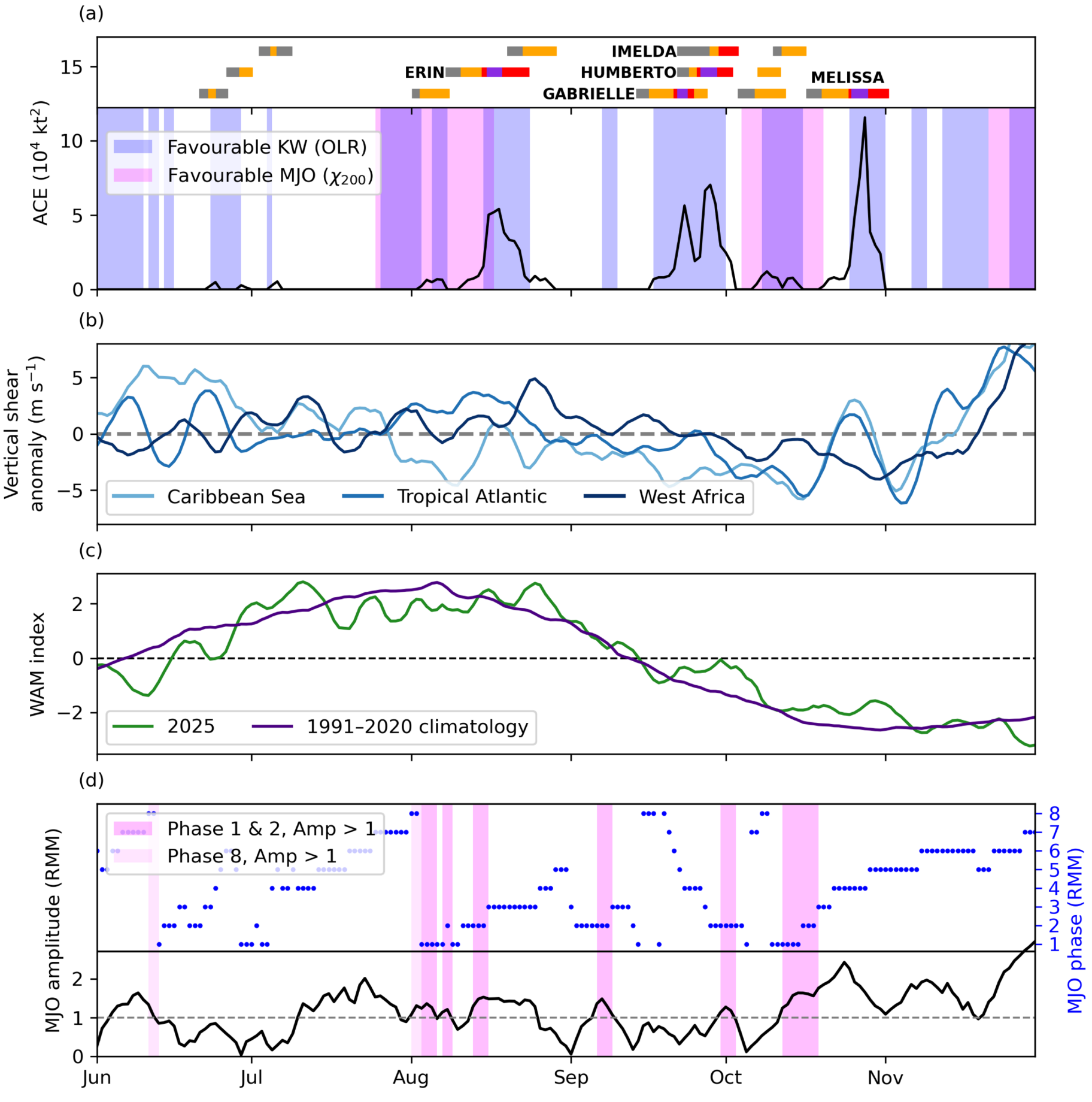

*Figure 6. Time series of activity, vertical shear, WAM circulation strength and favourable KW and MJO periods in the 2025 Atlantic hurricane season. (a) Daily ACE with periods of favourable MJO and KW activity highlighted in magenta and blue, respectively, corresponding to the presence of favourable wave signals as shown in Figure 5. Lifecycles of 2025 TCs shown with intensity indicated in colour as in Figure 3. (b) Daily mean 850–200hPa vertical wind shear anomaly in the Caribbean Sea, Tropical Atlantic and West Africa. Averaging regions shown in Figure 1. (c) WAM circulation index, defined as in Ndiaye* et al. *(2009) (see text). (d) MJO amplitude and phase calculated from the Real-time Multivariate MJO (RMM) index (Wheeler and Hendon, 2004) provided by the Australian Bureau of Meteorology.*

One of the strongest sources of intraseasonal variability in TC activity in the Atlantic basin is the MJO. Typically, the MJO favours TC formation in the Atlantic basin when in phases one and two, when upper-level divergence is enhanced over the African continent and Indian Ocean (strengthening the WAM and African easterly jet, influencing AEW development) and to a lesser extent also phase eight (Klotzbach, 2010; Barnston *et al*., 2015). The phase and amplitude of the MJO is often characterised using the RMM index (Wheeler and Hendon, 2004) shown in Figure 6(d); the MJO is considered 'coherent' when the RMM amplitude exceeds 1. Periods with a coherent and favourable MJO are highlighted, which are broadly consistent with the favourable periods identified via wave-filtering as shown in Figures 5 and 6(a). On average in 1991–2020, there were 49 days during the Atlantic hurricane season where the MJO was coherent and in phase 1, 2 or 8, with a standard deviation of 13 days. In 2025, there were only 25 days with these favourable MJO conditions, the second lowest since 1991. Combined with the increased stability and high SLP throughout the 2025 season, the relative lack of favourable MJO conditions may have contributed to the lack of TC activity relative to expectations based on the warm SSTs. It can be seen from Figures 5 and 6 that favourable MJO forcing occurred in two periods, coinciding with the development of TCs from AEWs: first, *Erin* and *Fernando* in mid-August, later, *Lorenzo* and *Melissa* in mid-October. The AEWs from which *Fernando*, *Lorenzo* and *Melissa* formed were also influenced by KWs.

## Circulation and SST patterns

The relative lack of activity in the 2025 season can be viewed as a result of generally unfavourable large-scale conditions (SLP, stability), as well as disrupted AEW development early (June, July) and late (November) in the season owing to smaller-scale variability in conditions (shear, humidity) and WAM strength. Here, we discuss how the upper-level circulation as well as low-level trade winds and SST patterns contributed to the observed intraseasonal variability. In boreal summer months, the upper-level circulation pattern in the North Atlantic features a TUTT adjacent to the subtropical (Azores) high. The strength, position and orientation of the TUTT are subject to internal variability as well as influence from mid-latitude wave-breaking and TCs (Ferreira and Schubert, 1999; Lu *et al*., 2017). The TUTT can also influence TC genesis (Sears and Velden, 2014) and intensification (Fischer *et al*., 2017). The upper-level flow brings cold air from the mid-latitudes into the subtropics, increasing lapse rates that favour convection, as well as enhancing divergence equatorward and east of the trough. However, the strong upper-level winds can also increase deep-layer shear and, depending on extent and position, mix dry subtropical air into the tropics and draw the SAL into the tropical Atlantic, suppressing TC activity. The movement of TCs is also heavily influenced by upper atmospheric winds that act as a 'steering flow', meaning a TUTT can promote recurvature in the mid-Atlantic.

Figure 7 shows monthly-mean upper-tropospheric winds and geopotential height contours in the Atlantic basin from June to September, with upper-level troughs indicated by blue lines, illustrating the persistent TUTT present in June and July that contributed to the strong vertical shear and dry tropical Atlantic noted earlier. Although the TUTT weakened in August, strong vertical shear was still present near West Africa that likely continued to disrupt AEW development. In September, the TUTT regained strength but with more favourable positioning and weaker shear that may have supported intensification of TCs moving westward across the Atlantic. Owing to the two-way feedback between TCs and TUTTs, the monthly averages in Figure 7 do not exclude days with large ACE values, unlike the averages shown in Figure 4. It is possible that the increase in TC activity in September contributed to variability in the TUTT: note that *Imelda* and *Humberto* formed between 40°W and 60°W, close to the trough axis.

It is notable that most TCs remained well offshore this season. The exceptions are Barry and Chantal, which made landfall in the US as tropical storms early in the season, and *Melissa*, which made landfall in Jamaica at peak strength as a Category 5 major hurricane and later in Cuba as a Category 3 hurricane. Although the TUTT varied in strength through the season, the upper-level circulation pattern favoured recurving TCs throughout. Persistent low SLP near the US coast in August–October also likely contributed to the recurvature of TCs (see Figure 4b).

Variability in SST patterns and low-level winds can influence TC activity by shifting convection patterns such as the ITCZ via changes in surface convergence and evaporation rates (Kossin and Vimont, 2007; Wang *et al*., 2025). Figure 8 shows the (a, b) 10m wind speed anomaly and (c, d) SST anomaly. Low-level 10m trade winds are overlaid. Days with ACE > 1 are excluded as in Figure 4. Surface convergence zones in 2025 and 1991–2020 climatology are shown using a simplified version of the Berry and Reeder (2014) methodology that identifies convex minima in the 925hPa convergence field. The dominant mode of interannual variability in tropical SSTs is the Atlantic equatorial mode (AEM), also referred to as the Atlantic zonal mode or Atlantic Niño (Niña) in the positive (negative) phase (Dippe *et al*., 2019). The AEM tends to develop in late boreal spring, peaking in May to July, and is closely related to the formation of a 'cold-tongue' of cooler-than-average SSTs in the eastern equatorial Atlantic during the negative phase (Keenlyside and Latif, 2007; Dippe *et al*., 2018). In the warm phase (Atlantic Niño), the AEM strengthens the ITCZ and WAM, increasing AEW activity and surface convergence in the tropical eastern Atlantic (Kim *et al*., 2023). Conversely, the presence of an equatorial cold-tongue during the cold phase (Atlantic

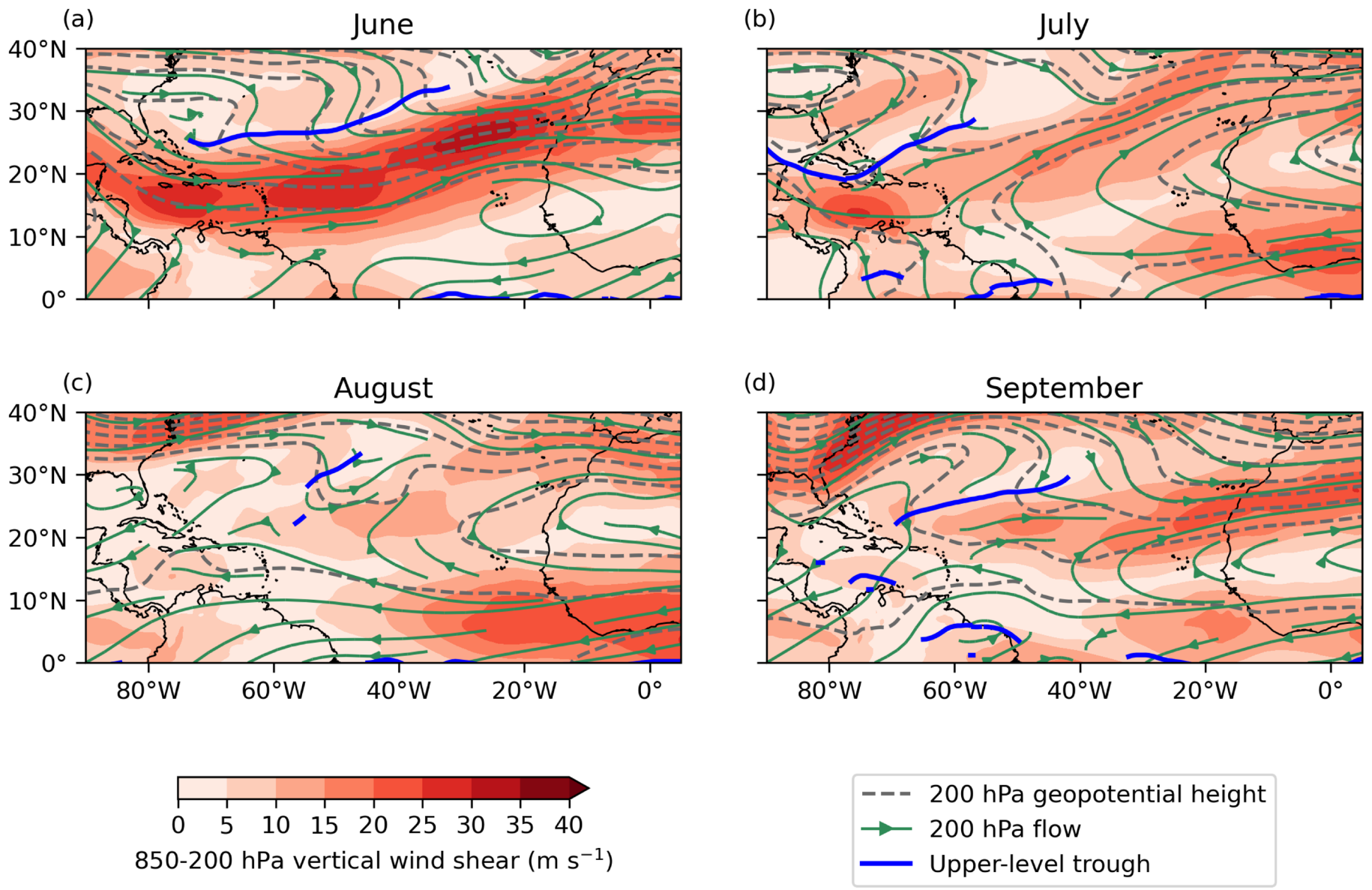

*Figure 7. Upper-troposphere (200hPa) geopotential height contours (grey dashed, 30m interval) and flow streamlines (green) with 850–200hPa vertical shear (coloured) in June to September 2025. Troughs are indicated by thick blue lines. Data from preliminary ERA5 reanalysis (Hersbach* et al.*, 2020).*

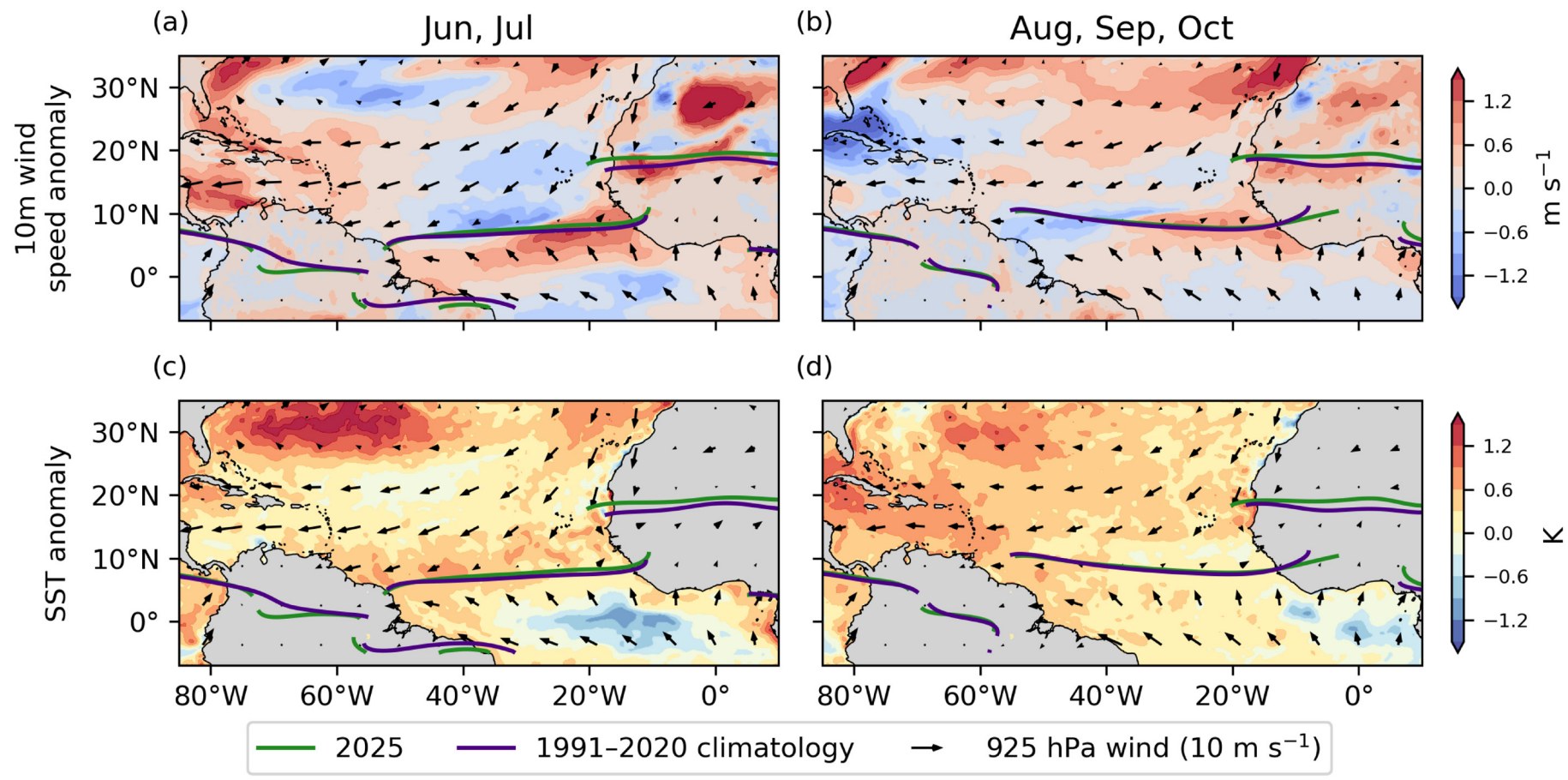

*Figure 8. Low-level (10m) winds overlaid on (a, b) 10m wind speed anomaly (colour) and (c, d) SST anomaly (colour) relative to 1991–2020 climatology averaged over (a, c) June & July and (b, d) August-October. Position of the ITCZ indicated by lines of convergence at 925hPa in 2025 (green) and 1991–2020 climatology (purple) using a modified version of the Berry and Reeder (2014) methodology. Data from preliminary ERA5 reanalysis (Hersbach* et al.*, 2020).*

Niña) enhances the temperature gradient between the Gulf of Guinea and the Sahel, strengthening low-level southerly winds that displace the WAM and African easterly jet (AEJ) northwards (Okumura and Xie, 2004; Caniaux *et al*., 2011). Changes in diabatic heating associated with a northward-shifted AEJ have been linked to weakened AEW activity (Núñez Ocasio *et al*., 2024, 2025). A negative phase of the AEM (Atlantic Niña, the first since 2014) developed in June, lasting until August and returning to neutral from September onwards (NOAA Climate Prediction Center, 2026a). The negative-AEM associated equatorial cold-tongue is clearly seen between 0°W and 30°W in Figure 8(c), consistent with a weakened WAM and disrupted AEW activity in June, July and much of August as noted in the earlier analysis of AEWs. During August–October, the cold-tongue is less pronounced (Figure 8d).

In June and July, the northeasterly trade winds between 60°W and 20°W were weak and the ITCZ was shifted slightly north. Although weakened trade winds typically weaken vertical shear, the anomalously strong upper-level winds counteracted this response (Figure 7). In August–October, the Atlantic segment of the ITCZ was close to its climatological position but the West African segment (associated with the WAM) was still displaced northwards, whilst the 60°–20°W trade winds were anomalously strong. The northward shifted convergence zone in West Africa likely resulted in a more northward track of AEWs. Stronger vertical wind shear (Figure 7) and cooler SSTs (Figure 8c, d) mean that AEWs on a northward shifted track tend to be weaker and therefore less likely to develop into TCs, as observed in 2024 (Klotzbach *et al*., 2025) and supported by the generally low AEW amplitudes noted in Figure 5. The increasing strength of the trade winds may have contributed to the cooler SSTs in the 5°–15°N belt by increasing surface evaporation rates from the early to late season. Similarly, the weakening trade winds in the Caribbean Sea from early to late season (compare Figure 8c and d) likely contributed to the increase in SSTs in the region (noted earlier and shown in Figure 4).

## Conclusion

The 2025 Atlantic hurricane season was near-normal in terms of cumulative ACE relative to the recent climatology period from 1991 onwards, despite weakly favourable La Niña conditions and above-average SSTs in the MDR that in previous seasons have led to above-average activity. The 2025 season was the second most intermittent of all seasons since 1991 with near-normal activity; the majority of ACE was produced by four major hurricanes out of a total of 13 named storms. In this sense, the 2025 season fits into the model-predicted trend of decreasing activity alongside increasing intensity of individual TCs (Knutson *et al*., 2020).

In this paper, we discussed the factors that influenced the 2025 season, identifying unfavourable large-scale conditions in the form of high SLP and increased atmospheric stability throughout. The early season (June and July) was quiet owing to a disrupted WAM, resulting in weak AEWs (diagnosed using wave-filtered relative vorticity) that were not able to develop in the Atlantic basin. Strong vertical shear and an extended dry air layer near West Africa from a persistent strong TUTT also disrupted AEWs during this period. Stronger AEWs moving into the Atlantic basin during August, likely supported by an increase in KW activity and favourable MJO conditions, led to an increase in activity with the first major hurricane of the season (*Erin*). During the climatological peak of the hurricane season, from mid-August to mid-September, no activity occurred. Vertical shear, unfavourable circulation and SST patterns from an Atlantic Niña event, and strong subsidence disrupted the development of AEWs and suppressed cyclogenesis. September and October saw two further clusters of activity as vertical shear weakened and the WAM regained strength. Hurricanes *Gabrielle*, *Humberto* and *Imelda* all formed and reached peak strength during a 2-week period in late September. The restrengthened and more favourably positioned TUTT, as well as more frequent KW activity, may have supported the development of AEWs crossing the Atlantic basin. The final burst of activity in October coincided with favourable MJO conditions, during which *Melissa* formed and later developed into an intense Category 5 hurricane owing to particularly warm SSTs in the Caribbean Sea that were aided by weakened trade winds in the preceding months.

Recent research showing that convectively active KW phases can modulate the development of AEWs is supported by the wave analysis presented in this paper (Lawton and Majumdar, 2023; Rios-Berrios *et al*., 2024b). This study represents an addition to recent case studies of KW influence on cyclogenesis using observations in the Atlantic (Ventrice *et al*., 2012a; Jonville *et al*., 2025a) and globally (Schreck, 2015, 2016). The extended periods of low activity in the 2025 season coincided with an absence of KW activity, particularly in July. Compared to recent climatology, there were also relatively few episodes of favourable MJO conditions. The factors that influence KW and MJO variability were not studied here but represent an important topic for future research, particularly their predictability. Further, it remains an open question whether KW influence on the development of AEWs and modulation of the large-scale environment by the MJO will become more important in a future climate with a more stable Atlantic. Other aspects of predictability that could be explored in future work include the persistence of TUTTs, which played a key role in controlling vertical wind shear in the 2025 season, and the interactions between different modes of variability in SST patterns such as the AEM which appeared to 'mask' the expected response to a positive Atlantic Meridional Mode in the latter part of the 2025 season.

Many interesting aspects of the 2025 season have not been covered in this study. One key aspect is rapid intensification (RI): in the North Atlantic, around 80% of TCs that undergo RI become major hurricanes, and there is evidence that RI is becoming more common (Bhatia *et al*., 2019). There were several examples of RI observed this season as the major hurricanes developed, which appears to support that trend. Another aspect is the continued advancement of AI tracking and intensity models, some of which appeared to pick up on episodes of RI ahead of conventional physical models (NHC, 2025).

## Acknowledgements

This project is supported by Inigo Limited. The author is grateful for insightful comments from Alison Ming, Ruth Petrie, Sebastian Schemm and two anonymous reviewers during preparation of this manuscript.

## Data availability statement

Scripts used to generate the figures in this manuscript can be found at https://github.com/qntmCharles/retro_2025. The following datasets have been used: Optimum Interpolation SST (OISSTv2.1), available from the National Oceanic and Atmospheric Administraiton (NOAA) (https://www.ncei.noaa.gov/products/optimum-interpolation-sst). Atlantic Hurricane Database (HURDAT2), available from the National Hurricane Center (NHC) (https://www.nhc.noaa.gov/data/hurdat/). Automated Tropical Cyclone Forecast (ATCF) system best-track (btk) files, available from the NHC FTP server (https://ftp.nhc.noaa.gov/atcf/btk/). Fifth generation of ECMWF atmospheric reanalyses of the global climate (ERA, preliminary), obtained from the Copernicus Climate Change Service (C3S) (https://cds.climate.copernicus.eu). Climate Forecast System (CFS) reanalysis, available from the National Centre for Atmospheric Research (NCAR) (https://climatedataguide.ucar.edu/climate-data/climate-forecast-system-reanalysis-cfsr). MJO RMM index data provided by the Australian Bureau of Meteorology (https://www.bom.gov.au/climate/mjo/). NOAA Physical Sciences Laboratory (PSL) (https://psl.noaa.gov/).

## References

**Afiesimama EA**. 2007. Annual cycle of the mid-tropospheric easterly jet over West Africa. *Theor. Appl. Climatol.* **90**(1–2): 103–111. https://doi.org/10.1007/s00704-006-0284-y

**Agudelo PA**, **Hoyos CD**, **Curry JA** *et al*. 2011. Probabilistic discrimination between large-scale environments of intensifying and decaying African easterly waves. *Clim. Dyn.* **36**(7–8): 1379–1401. https://doi.org/10.1007/s00382-010-0851-x

**Barnston AG**, **Vigaud N**, **Long LN** *et al*. 2015. Atlantic tropical cyclone activity in response to the MJO in NOAA's CFS model. *Mon. Weather Rev.* **143**(12): 4905–4927. https://doi.org/10.1175/MWR-D-15-0127.1

**Barrett BS**, **Leslie LM**. 2009. Links between tropical cyclone activity and Madden–Julian oscillation phase in the North Atlantic and northeast Pacific basins. *Mon. Weather Rev.* **137**(2): 727–744. https://doi.org/10.1175/2008MWR2602.1

**Berry G**, **Reeder MJ**. 2014. Objective identification of the intertropical convergence zone: climatology and trends from the ERA-interim. *J. Clim.* **27**(5): 1894–1909. https://doi.org/10.1175/JCLI-D-13-00339.1

**Bhatia KT**, **Vecchi GA**, **Knutson TR** *et al*. 2019. Recent increases in tropical cyclone intensification rates. *Nat. Commun.* **10**(1): 635. https://doi.org/10.1038/s41467-019-08471-z

**Caniaux G**, **Giordani H**, **Redelsperger J-L** *et al*. 2011. Coupling between the Atlantic cold tongue and the West African monsoon in boreal spring and summer. *J. Geophys. Res.* **116**(C4): C04003. https://doi.org/10.1029/2010JC006570

**Corporal-Lodangco IL**, **Richman MB**, **Leslie LM** *et al*. 2014. Cluster analysis of North Atlantic tropical cyclones. *Procedia Comput. Sci.* **36**: 293–300. https://doi.org/10.1016/j.procs.2014.09.096

**Dippe T**, **Greatbatch RJ**, **Ding H**. 2018. On the relationship between Atlantic Niño variability and ocean dynamics. *Clim. Dyn.* **51**(1–2): 597–612. https://doi.org/10.1007/s00382-017-3943-z

**Dippe T**, **Greatbatch RJ**, **Ding H**. 2019. Seasonal prediction of equatorial Atlantic sea surface temperature using simple initialization and bias correction techniques. *Atmos. Sci. Lett.* **20**(5): e898. https://doi.org/10.1002/asl.898

**Emanuel K**. 2003. Tropical cyclones. *Annu. Rev. Earth Planet. Sci.* **31**(1): 75–104. https://doi.org/10.1146/annurev.earth.31.100901.141259

**Emanuel K**. 2005. Increasing destructiveness of tropical cyclones over the past 30 years. *Nature* **436**(7051): 686–688.

**Ferreira RN**, **Schubert WH**. 1999. The role of tropical cyclones in the formation of tropical upper-tropospheric troughs. *J. Atmos. Sci.* **56**(16): 2891–2907. https://doi.org/10.1175/1520-0469(1999)056⟨2891:TROTCI⟩2.0.CO;2

**Fischer MS**, **Tang BH**, **Corbosiero KL**. 2017. Assessing the influence of upper-tropospheric troughs on tropical cyclone intensification rates after genesis. *Mon. Weather Rev.* **145**(4): 1295–1313. https://doi.org/10.1175/MWR-D-16-0275.1

**Fitzpatrick RGJ**, **Bain CL**, **Knippertz P** *et al*. 2015. The West African monsoon onset: a concise comparison of definitions. *J. Clim.* **28**(22): 8673–8694. https://doi.org/10.1175/JCLI-D-15-0265.1

**Goldenberg SB**, **Landsea CW**, **Mestas-Nuñez AM** *et al*. 2001. The recent increase in Atlantic hurricane activity: causes and implications. *Science* **293**(5529): 474–479. https://doi.org/10.1126/science.1060040

**Gray WM**. 1968. Global view of the origin of tropical disturbances and storms. *Mon. Weather Rev.* **96**(10): 669–700. https://doi.org/10.1175/1520-0493(1968)096⟨0669:GVOTOO⟩2.0.CO;2

**Gray WM**. 1984. Atlantic seasonal hurricane frequency. Part I: El Niño and 30 mb quasi-biennial oscillation influences. *Mon. Weather Rev.* **112**(9): 1649–1668. https://doi.org/10.1175/1520-0493(1984)112⟨1649:ASHFPI⟩2.0.CO

**Hersbach H**, **Bell B**, **Berrisford P** *et al*. 2020. The ERA5 global reanalysis. *Q. J. R. Meteorol. Soc.* **146**(730): 1999–2049. https://doi.org/10.1002/qj.3803

**Hopsch SB**, **Thorncroft CD**, **Tyle KR**. 2010. Analysis of African easterly wave structures and their role in influencing tropical cyclogenesis. *Mon. Weather Rev.* **138**(4): 1399–1419. https://doi.org/10.1175/2009MWR2760.1

**Hsieh J-S**, **Cook KH**. 2005. Generation of African easterly wave disturbances: relationship to the African easterly jet. *Mon. Weather Rev.* **133**(5): 1311–1327. https://doi.org/10.1175/MWR2916.1

**Huang B**, **Liu C**, **Banzon V** *et al*. 2021. Improvements of the daily optimum interpolation sea surface temperature (DOISST) version 2.1. *J. Clim.* **34**(8): 2923–2939. https://doi.org/10.1175/JCLI-D-20-0166.1

**Jonville T**, **Borne M**, **Flamant C** *et al*. 2025a. Impact of convectively coupled tropical waves on the composition, vertical structure of the atmosphere, and tropical cyclogenesis in the region of Cabo Verde in September 2021 during the CADDIWA campaign. *Atmos. Chem. Phys.* **25**(17): 9765–9786. https://doi.org/10.5194/acp-25-9765-2025

**Jonville T**, **Cornillault E**, **Lavaysse C** *et al*. 2025b. Distinguishing north and south African Easterly Waves with a spectral method: implication for tropical cyclogenesis from mergers in the North Atlantic. *Q. J. R. Meteorol. Soc.* **151**(767): e4909. https://doi.org/10.1002/qj.4909

**Karnauskas KB**, **Zhang L**, **Emanuel KA**. 2021. The feedback of cold wakes on tropical cyclones. *Geophys. Res. Lett.* **48**(7): e2020GL091676. https://doi.org/10.1029/2020GL091676

**Keenlyside NS**, **Latif M**. 2007. Understanding equatorial Atlantic interannual variability. *J. Clim.* **20**(1): 131–142. https://doi.org/10.1175/JCLI3992.1

**Keil P**, **Schmidt H**, **Stevens B** *et al*. 2021. Variations of tropical lapse rates in climate models and their implications for upper tropospheric warming. *J. Clim.* 1–50. https://doi.org/10.1175/JCLI-D-21-0196.1

**Kim D**, **Lee S-K**, **Lopez H** *et al*. 2023. Increase in Cape Verde hurricanes during Atlantic Niño. *Nat. Commun.* **14**(1): 3704. https://doi.org/10.1038/s41467-023-39467-5

**Klotzbach PJ**. 2010. On the Madden–Julian oscillation–Atlantic hurricane relationship. *J. Clim.* **23**(2): 282–293. https://doi.org/10.1175/2009JCLI2978.1

**Klotzbach PJ**. 2011. El Niño–southern oscillation's impact on Atlantic basin hurricanes and U.S. landfalls. *J. Clim.* **24**(4): 1252–1263. https://doi.org/10.1175/2010JCLI3799.1

**Klotzbach PJ**, **Bercos-Hickey E**, **Wood KM** *et al*. 2025. The remarkable 2024 North Atlantic mid-season hurricane lull. *Geophys. Res. Lett.* **52**(19): e2025GL116714. https://doi.org/10.1029/2025GL116714

**Klotzbach PJ**, **Gray WM**. 2008. Multidecadal variability in North Atlantic tropical cyclone activity. *J. Clim.* **21**(15): 3929–3935.

**Klotzbach PJ**, **Oliver ECJ**. 2015. Variations in global tropical cyclone activity and the Madden-Julian oscillation since the midtwentieth century. *Geophys. Res. Lett.* **42**(10): 4199–4207. https://doi.org/10.1002/2015GL063966

**Knaff JA**. 1997. Implications of summertime sea level pressure anomalies in the tropical Atlantic region. *J. Clim.* **10**(4): 789–804. https://doi.org/10.1175/1520-0442(1997)010⟨0789:IOSSLP⟩2.0.CO;2

**Knutson T**, **Camargo SJ**, **Chan JCL** *et al.* 2020. Tropical cyclones and climate change assessment: part II: projected response to anthropogenic warming. *Bull. Am. Meteorol. Soc.* **101**(3): E303–E322. https://doi.org/10.1175/BAMS-D-18-0194.1

**Kossin JP**, **Vimont DJ**. 2007. A more general framework for understanding Atlantic hurricane variability and trends. *Bull. Am. Meteorol. Soc.* **88**(11): 1767–1782. https://doi.org/10.1175/BAMS-88-11-1767

**Landsea CW**, **Franklin JL**. 2013. Atlantic hurricane database uncertainty and presentation of a new database format. *Mon. Weather Rev.* **141**(10): 3576–3592. https://doi.org/10.1175/MWR-D-12-00254.1

**Lawton QA**, **Majumdar SJ**. 2023. Convectively coupled Kelvin waves and tropical cyclogenesis: connections through convection and moisture. *Mon. Weather Rev.* **151**(7): 1647–1666. https://doi.org/10.1175/MWR-D-23-0005.1

**Lawton QA**, **Majumdar SJ**, **Dotterer K** *et al.* 2022. The influence of convectively coupled Kelvin waves on African easterly waves in a wave-following framework. *Mon. Weather Rev.* **150**(8): 2055–2072. https://doi.org/10.1175/MWR-D-21-0321.1

**Lu M**, **Deng K**, **Yang S** *et al.* 2017. Interannual and interdecadal variations of the Mid-Atlantic trough and associated American-Atlantic-Eurasian climate anomalies. *Atmos.-Ocean* **55**(4–5): 284–292. https://doi.org/10.1080/07055900.2017.1369931

**Maloney ED**, **Hartmann DL**. 2000. Modulation of hurricane activity in the Gulf of Mexico by the Madden-Julian oscillation. *Science* **287**(5460): 2002–2004. https://doi.org/10.1126/science.287.5460.2002

**McTaggart-Cowan R**, **Davies EL**, **Fairman JG** *et al.* 2015. Revisiting the 26.5°C sea surface temperature threshold for tropical cyclone development. *Bull. Am. Meteorol. Soc.* **96**(11): 1929–1943. https://doi.org/10.1175/BAMS-D-13-00254.1

**Ndiaye O**, **Goddard L**, **Ward MN**. 2009. Using regional wind fields to improve general circulation model forecasts of July–September Sahel rainfall. *Int. J. Climatol.* **29**(9): 1262–1275. https://doi.org/10.1002/joc.1767

**NHC**. 2025. Tropical Storm Melissa Discussion Number 10. https://www.nhc.noaa.gov/archive/2025/al13/al132025.discus.010.shtml (accessed 28 November 2025).

**NHC**. 2026. Tropical cyclone reports. https://www.nhc.noaa.gov/data/tcr/index.php (accessed 11 February 2026).

**NOAA**. 2025. NOAA ENSO diagnostic discussion. https://www.cpc.ncep.noaa.gov/products/analysis_monitoring/enso_advisory/ (accessed 28 November 2025).

**NOAA**. 2026. Destructive 2018 Atlantic Hurricane season draws to an end. https://www.noaa.gov/media-release/destructive-2018-atlantic-hurricane-season-draws-to-end (accessed 17 February 2026).

**NOAA Climate Prediction Center**. 2026a. Atlantic 3 Index. https://www.cpc.ncep.noaa.gov/products/international/ocean_monitoring/IODMI/ATL3_month.html (accessed 13 February 2026).

**NOAA Climate Prediction Center**. 2026b. Relative Oceanic Nino Index. https://www.cpc.ncep.noaa.gov/products/analysis_monitoring/enso/roni/ (accessed 17 February 2026).

**Núñez Ocasio KM**, **Davis CA**, **Moon ZL** *et al.* 2024. Moisture dependence of an African easterly wave within the West African monsoon system. *J. Adv. Model. Earth Syst.* **16**(6): e2023MS004070.

**Núñez Ocasio KMN**, **Dougherty EM**, **Moon ZL** *et al.* 2025. Response of African easterly waves to a warming climate: a convection-permitting approach. *J. Adv. Model. Earth Syst.* **17**(10): e2025MS005146.

**Okumura Y**, **Xie S-P**. 2004. Interaction of the Atlantic equatorial cold tongue and the African monsoon. *J. Clim.* **17**(18): 3589–3602. https://doi.org/10.1175/1520-0442(2004)017⟨3589:IOTAEC⟩2.0.CO;2

**Pu B**, **Cook KH**. 2012. Role of the West African westerly jet in Sahel rainfall variations. *J. Clim.* **25**(8): 2880–2896. https://doi.org/10.1175/JCLI-D-11-00394.1

**Rajasree V**, **Cao X**, **Ramsay H** *et al.* 2023. Tropical cyclogenesis: controlling factors and physical mechanisms. *Trop. Cyclone Res. Rev.* **12**(3): 165–181. https://doi.org/10.1016/j.tcrr.2023.09.004

**Rios-Berrios R**, **Finocchio PM**, **Alland JJ** *et al.* 2024a. A review of the interactions between tropical cyclones and environmental vertical wind shear. *J. Atmos. Sci.* **81**(4): 713–741. https://doi.org/10.1175/JAS-D-23-0022.1

**Rios-Berrios R**, **Tang BH**, **Davis CA** *et al.* 2024b. Modulation of tropical cyclogenesis by convectively coupled Kelvin waves. *Mon. Weather Rev.* **152**(10): 2309–2322. https://doi.org/10.1175/MWR-D-24-0052.1

**Russell JO**, **Aiyyer A**, **White JD** *et al.* 2017. Revisiting the connection between African easterly waves and Atlantic tropical cyclogenesis. *Geophys. Res. Lett.* **44**(1): 587–595. https://doi.org/10.1002/2016GL071236

**Sampson CR**, **Schrader AJ**. 2000. The automated tropical cyclone forecasting system (Version 3.2). *Bull. Am. Meteorol. Soc.* **81**(6): 1231–1240. https://doi.org/10.1175/1520-0477(2000)081⟨1231:TATCFS⟩2.3.CO;2

**Schneider DP**, **Deser C**, **Fasullo J** *et al.* 2013. Climate data guide spurs discovery and understanding. *EoS Trans.* **94**(13): 121–122. https://doi.org/10.1002/2013EO130001

**Schreck CJ**. 2015. Kelvin waves and tropical cyclogenesis: a global survey. *Mon. Weather Rev.* **143**(10): 3996–4011.

**Schreck CJ**. 2016. Convectively coupled Kelvin waves and tropical cyclogenesis in a semi-Lagrangian framework. *Mon. Weather Rev.* **144**(11): 4131–4139. https://doi.org/10.1175/MWR-D-16-0237.1

**Schreck CJ**, **Molinari J**, **Aiyyer A**. 2012. A global view of equatorial waves and tropical cyclogenesis. *Mon. Weather Rev.* **140**(3): 774–788. https://doi.org/10.1175/MWR-D-11-00110.1

**Sears J**, **Velden CS**. 2014. Investigating the role of the upper-levels in tropical cyclone genesis. *Trop. Cyclone Res. Rev.* **3**(2): 91–110.

**Thorncroft C**, **Hodges K**. 2001. African easterly wave variability and its relationship to Atlantic tropical cyclone activity. *J. Clim.* **14**(6): 1166–1179. https://doi.org/10.1175/1520-0442(2001)014⟨1166:AEWVAI⟩2.0.CO;2

**Trenberth KE**, **Shea DJ**. 2006. Atlantic hurricanes and natural variability in 2005. *Geophys. Res. Lett.* **33**(12): 2006GL026894. https://doi.org/10.1029/2006GL026894

**UCAR**. 2025. Record-breaking winds confirmed for Hurricane Melissa. https://news.ucar.edu/133047/record-breaking-winds-confirmed-hurricane-melissa (accessed 28 November 2025).

**Vellinga M**, **Arribas A**, **Graham R**. 2013. Seasonal forecasts for regional onset of the West African monsoon. *Clim. Dyn.* **40**(11–12): 3047–3070. https://doi.org/10.1007/s00382-012-1520-z

**Ventrice MJ**, **Thorncroft CD**, **Janiga MA**. 2012a. Atlantic tropical cyclogenesis: a three-way interaction between an African easterly wave, diurnally varying convection, and a convectively coupled atmospheric Kelvin wave. *Mon. Weather Rev.* **140**(4): 1108–1124. https://doi.org/10.1175/MWR-D-11-00122.1

**Ventrice MJ**, **Thorncroft CD**, **Schreck CJ**. 2012b. Impacts of convectively coupled Kelvin waves on environmental conditions for Atlantic tropical cyclogenesis. *Mon. Weather Rev.* **140**(7): 2198–2214. https://doi.org/10.1175/MWR-D-11-00305.1

**Wang Y**, **Satoh M**, **Zhan R** *et al.* 2025. Tropical sea surface warming patterns and tropical cyclone activity: a review. *Adv. Atmos. Sci.* **42**(10): 1996–2017. https://doi.org/10.1007/s00376-025-5114-1

**Wheeler MC**, **Hendon HH**. 2004. An all-season real-time multivariate MJO index: development of an index for monitoring and prediction. *Mon. Weather Rev.* **132**(8): 1917–1932. https://doi.org/10.1175/1520-0493(2004)132⟨1917:AARMMI⟩2.0.CO;2

**Wilks DS**. 2006. Statistical Methods in the Atmospheric Sciences, 2nd Edition, Vol. **91**. Elsevier: Amsterdam.

**Williams AIL**, **Jeevanjee N**, **Bloch-Johnson J**. 2023. Circus tents, convective thresholds, and the non-linear climate response to tropical SSTs. *Geophys. Res. Lett.* **50**(6): e2022GL101499. https://doi.org/10.1029/2022GL101499

**Zhang G**, **Cook KH**. 2014. West African monsoon demise: climatology, interannual variations, and relationship to seasonal rainfall. *J. Geophys. Res. Atmos.* **119**(17). https://doi.org/10.1002/2014JD022043

Correspondence to: C. W. Powell
*cwp29@cam.ac.uk*